\documentclass[12pt]{iopart}
\usepackage{amssymb}
\usepackage{color}
\usepackage{graphicx}
\usepackage{bm}
\usepackage{subfigure}
\usepackage{enumitem}
\usepackage{epstopdf}

\newcommand{\diag}{\mathop{\mathrm{diag}}}

\definecolor{dgreen}{rgb}{0,0.5,0}

\definecolor{delete}{cmyk}{0.5,0,0,0}

\newcommand{\eq}{Eq.~}
\newcommand{\eqs}{Eqs.~}

\newcommand{\cf} {cf.~}

\newcommand{\eg} {e.g.~}

\newcommand{\rref} {Ref.~}
\newcommand{\rrefs} {Refs.~}

\begin{document}
\def\bbm[#1]{\mbox{\boldmath$#1$}}

\title{Selective writing and read-out of a register of static qubits}

\author{A. De Pasquale}
\address{NEST, Scuola Normale Superiore and Istituto di Nanoscienze-CNR, 
Piazza dei Cavalieri 7, 
I-56126 Pisa, Italy}

\author{F. Ciccarello}
\address{NEST, Scuola Normale Superiore and Istituto di Nanoscienze-CNR, 
Piazza dei Cavalieri 7, 
I-56126 Pisa, Italy}    
\address{NEST, Istituto Nanoscienze-CNR and Dipartimento di Fisica e Chimica, Universit\`a  degli Studi di Palermo, via Archirafi 36, I-90123 Palermo, Italy}

\author{K. Yuasa}
\address{Department of Physics, Waseda University, Tokyo 169-8555, Japan}  

\author{V. Giovannetti}
\address{NEST, Scuola Normale Superiore and Istituto di Nanoscienze-CNR, 
Piazza dei Cavalieri 7, 
I-56126 Pisa, Italy}

\begin{abstract}
We propose a setup comprising an arbitrarily large array of static qubits (SQs), which interact with a flying qubit (FQ). The SQs work as a quantum register, which can be written or read-out by means of the FQ through quantum state transfer (QST). The entire system, including the FQ's motional degrees of freedom, behaves quantum mechanically.
We demonstrate a strategy allowing for selective QST between the FQ and a single SQ chosen from the register. This is achieved through a perfect mirror located beyond the SQs and suitable modulation of the inter-SQ distances.
\end{abstract}

\maketitle

\section{Introduction}
\begin{figure}[b]
\begin{center}
\includegraphics[width=0.45\linewidth]{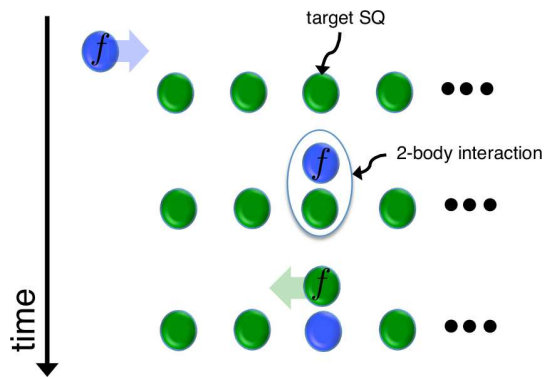}\end{center}
\caption{(Color online) Selective quantum state transfer between a FQ and a register of SQs. The FQ reaches the target SQ, exchanges its information content with it and eventually leaves the register.}
\label{fig1}
\end{figure}

A prominent paradigm in quantum information processing (QIP) \cite{nielsen} is to employ flying qubits (FQs) and static qubits (SQs) as carriers and registers of quantum information, respectively \cite{kimble}. Key to such idea is the ability to write and read-out the information content of a SQ by means of a FQ. By this, here we mean that efficient quantum state transfer (QST) between these two types of qubits must be possible on demand. 
In this picture, \textit{control over memory allocation} appears a desirable if not indispensable requirement. For instance,  one can envisage the situation where only one or a few SQs are available, \eg because the remaining ones are encoding some information to save. On the other hand, one may need to carry away only the information saved in certain specific SQs. Alternatively, only a restricted area of the register of SQs may be interfaced with some external processing network where one would like to eventually convey information or from which output data are to be received.
In such cases, the ability of \textit{selecting} the exact location where the information content of the FQ should be uploaded or downloaded is demanded. {Ideally, according to the schematics in Fig.~\ref{fig1}, 
one} would like the FQ to reach the specific target SQ, then fully transfer its quantum state to this and eventually fly away. 
Evidently, this picture is implicitly based on the assumption that, firstly, the motional degrees of freedom of the FQ are in fact fully classical and, secondly, that these can be accurately controlled. Despite its simplicity, although interesting research along this line is being carried out mostly through so called surface acoustic waves (see \eg \rref\cite{saw} and references therein), such an approach calls for a very high level of control.  

If we set within a fully quantum framework, the most natural situation to envisage is the one where the FQ, besides bearing an internal spin, moves in a quantum mechanical way and hence propagates as a wavelike object. Such a circumstance substantially complicates the dynamics in that, besides the complex spin-spin interactions, intricate wavelike effects such as multiple reflections between the many SQs occur either. This appears an adverse environment to accomplish \textit{selective} QST: while ideally one would like to focus the FQ's wave packet right on the target SQ, the former is expected to spread throughout the SQs' register. Thereby, 
not only  it is non-trivial what strategy would enable selective QST but even the mere possibility that this could occur can be questioned.

\begin{figure}[b]
\begin{center}
\includegraphics[width=0.55\linewidth]{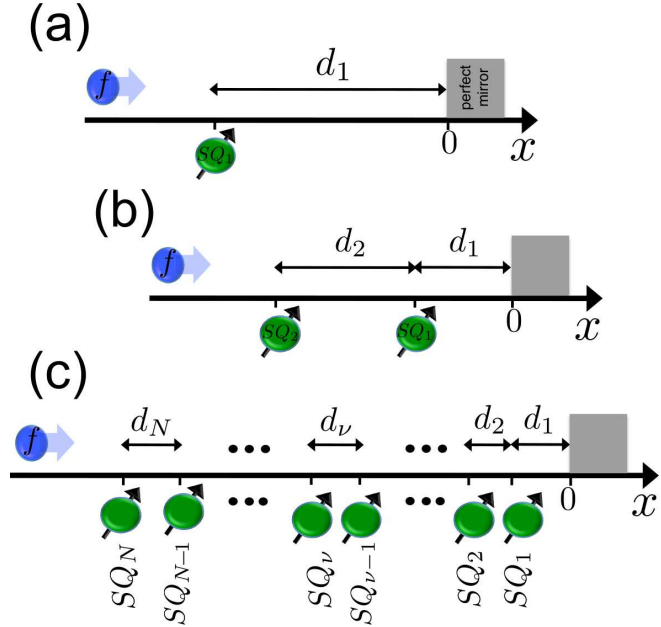}\end{center}
\caption{(Color online) Sketch of the setup in the case of one (a), two (b) and arbitrary $N$ (c) SQs. The FQ $f$ incomes from the left with a given wave vector $k$, undergoes multiple scattering between the SQs and the perfect mirror and eventually moves away from the register with the same $k$.}
\label{fig2}
\end{figure} 
In this work, we consider a paradigmatic Hamiltonian memory read-out model {where the FQ propagates along a 1D line comprising a collection of (fixed) spatially-separated non-interacting SQs and couples to them} via a contact-type spin-spin Heisenberg interactions (see Fig.~\ref{fig2}). We start with a single SQ and prove that a unitary \textsc{swap} between the itinerant and static spins is unattainable.
The insertion of a perfect mirror {along the 1D line}, however, makes it possible. At the same time, since the transmission channel is suppressed there is no uncertainty over the final path followed by the FQ\@. 
Next, we find that even for a pair of SQs this can be achieved with either of the two SQs through an \textit{ad hoc} setting of distances and coupling strengths. Surprisingly enough, this means that Feynman paths entering multiple reflections can combine so as to effectively decouple one SQ while enabling at the same time a unitary {\textsc{swap}} involving the other one.  Even more surprisingly, the working principle behind this phenomenon is such that it is naturally generalized to the case of an arbitrarily large register of SQs, as we rigorously prove.

\section{Read-out of a single static memory qubit}
Consider the  case  where a single memory static qubit  $SQ_1$ lies on the $x$-axis close to position $x= 0$. To read-out the quantum information stored in $SQ_1$ (or write it there) a FQ $f$ is injected along the axis with momentum $k$, say from the left-hand side.
We model the $f$-{$SQ_1$} interaction as a contact-type spin-dependent scattering potential having the Heisenberg coupling form. The system Hamiltonian can thus be expressed as $\hat{H}= \hat{p}^2/2 + \hat{V}$, where $\hat{p}$ is the momentum operator of $f$ (its mass being set equal to one for simplicity) and
\begin{equation}\label{V}
\hat{V}= G(\hat{\bm \sigma}_f\cdot\bm\hat{\bm \sigma}_{1})\delta(x)
\end{equation} 
 {is the coupling potential with associated strength $G$\footnote{{The assumption of {the} $\delta$-{shaped} potential is a standard one, and for the present setup it relies on the usually met condition that the FQ's wavelength is significantly larger than the characteristic SQ size}.}. Here, $x$ is the spatial coordinate of $f$ while  $\hat{\bm \sigma}_{f}$ and $\hat{\bm \sigma}_{1}$ are the spin operators
 of qubits $f$ and $SQ_1$, respectively, i.e.~$\hat{\bm \sigma}= (\sigma_x, \sigma_y, \sigma_z)$ with $\hat{\sigma}_{\beta=x,y,z}$ having eigenvalues $\pm 1/2$ (we set $\hbar= 1$ throughout). 
We ask whether or not,  when $f$ will emerge from the scattering process,  the internal degree of freedom (i.e.~the spin) of the two qubits have been exchanged according to the mapping }
 \begin{equation}\label{swapgate}
\rho_{f1} \; \rightarrow \; \rho^{\scriptsize(\textsc{swap})}_{f1} =   \hat{W}_{f1}  \rho_{f1}   \hat{W}_{f1}^\dag,
 \end{equation} 
where $\rho_{f1}$ 
is  the (joint) input spin state of $f$ and $SQ_1$, while $\hat{W}_{ij}$ is the usual \textsc{swap} two-qubit {unitary} operator exchanging the states of qubits $i$ and $j$ \cite{nielsen}. 
While there are in fact counterexamples \cite{amjphys, marsiglio} showing that this is impossible\footnote{In \rref\cite{marsiglio}, it was proven that, given the initial spin state $|{\uparrow\downarrow}\rangle_{f1}$, the scattering process between $f$ and $SQ_1$ can never lead to $\langle \hat{\sigma}_{1z}\rangle= 1/2$. Owing to conservation of $\hat{\sigma}_{fz}+ \hat{\sigma}_{1z}$, this is equivalent to state that the transformation $|{\uparrow\downarrow}\rangle_{f1}\rightarrow|{\downarrow\uparrow}\rangle_{f1}$ is unattainable.}, we give next the general proof that such \textsc{swap} operation cannot occur. 
For this purpose, let us define $|\Psi^{\pm}\rangle_{f1}= (|{\uparrow\downarrow}\rangle_{f1}\pm|{\downarrow\uparrow}\rangle_{f1})/\sqrt{2}$, where for each qubit, either flying or static, $|{\uparrow}\rangle$ and $|{\downarrow}\rangle$ stand for the eigenstates of $\hat{\sigma}_z$ with eigenvalues $1/2$ and $-1/2$, respectively (from now on, we omit particle subscripts whenever unnecessary). State $|\Psi^{-}\rangle$ is the well-known singlet, while the triplet subspace is spanned by $\{|{\uparrow\uparrow}\rangle, |\Psi^+\rangle, |{\downarrow\downarrow}\rangle\}$. 
Using the identity $\hat{\bm \sigma}_f \cdot \bm\hat{\bm \sigma}_{1}=  (\hat{\bm S}_{f1}^2-  \hat{\bm \sigma}_{f}^2-  \hat{\bm \sigma}_{1}^2)/2$, where $\hat{\bm S}_{f1}= \hat{\bm\sigma}_{f}+  \hat{\bm\sigma}_{1}$, the interaction Hamiltonian can be written as $\hat{V}=  (G/2) (\hat{\bm S}_{f1}^2-  3/2)\delta(x)$, entailing $[\hat{H},\hat{\bm S}_{f1}^2]= 0$ \cite{marsiglio,tomo,toment}. 
Within the singlet (triplet) subspace the effective interaction is thus \textit{spinless} and reads $\hat{V}_{\mathrm{s}}=  -(3G/4)  \delta(x)$ [$\hat{V}_{\mathrm {t}}=  (G/4) \delta(x)$]: the problem is reduced to a scattering from a (spin-independent) $\delta$-barrier. 
For a  $\delta$-potential step $\Gamma\delta(x)$ and a particle incoming with momentum $k$, the reflection and transmission probability amplitudes $r^{(0)}(\gamma)$ and $t^{(0)}(\gamma)$, respectively, are found through a textbook calculation as
\begin{equation}
r^{(0)}(\gamma)= t^{(0)}(\gamma) -  1=-i\gamma/(1+i\gamma),
\end{equation}
where we have introduced the rescaled parameter $\gamma=  \Gamma/k$.
These functions allow to calculate the reflection coefficient for the singlet and  triplet sectors  as 
\begin{eqnarray}
{r_{\rm s}=  t_{\rm s}-1=  r^{(0)}(-3g/4)} \quad (\textrm{singlet}), \label{DEFRandT1}\\ 
{r_{\rm t}=  t_{\rm t}-1=  r^{(0)}(g/4)} \qquad\,(\textrm{triplet}), \label{DEFRandT2}
\end{eqnarray}
where we have set $ g=  G/k$. 
Evidently, $|r_{\rm t}|\neq|r_{\rm s}|$ for any $G\neq0$. 
This is the very reason which forbids one from
using the above scattering  process for implementing \textit{any} unitary gate on the spin degree of freedom of $f$ and $SQ_1$, hence, in particular, the \textsc{swap} gate (\ref{swapgate}) enabling perfect writing/read-out of $SQ_1$. 
Observe in fact that, once the orbital degree of freedom of the FQ are traced out, the final spin state $\rho'_{f1}$ of the joint system $f$-$SQ_1$ can be related to the initial one $\rho_{f1}$ (in general mixed) through the completely positive, trace-preserving map \cite{nielsen} 
\begin{equation}
\rho_{f1} \rightarrow \rho_{f1}'=  \hat{T}_{f1}\rho_{f1}\hat{T}_{f1}^\dag+  \hat{R}_{f1}\rho_{f1}\hat{R}_{f1}^\dag,\label{mapping1}
\end{equation}
where the first contribution refers to the $f$-wave component emerging from the right of the 1D line (transmission channel), while the second to the one emerging from the left (reflection channel).
The Kraus operators \cite{nielsen,zyczkowski}  $\hat {T}_{f1}$ and $\hat {R}_{f1}$ describing these two complementary events  are provided,  
respectively, by the transmission  and reflection operators of the  model, namely
 \begin{equation}
 \hat {R}_{f1}= r_{\rm s} \hat{\Pi}_{f1}^{({\rm s})} + r_{\rm t} \hat{\Pi}_{f1}^{({\rm t})} , \qquad 
 \hat {T}_{f1} = t_{\rm s}\hat{\Pi}_{f1}^{({\rm s})} + t_{\rm t}\hat{\Pi}_{f1}^{({\rm t})}, \label{defrt}
 \end{equation} 
where  $\hat{\Pi}_{f1}^{({\rm s})}= |\Psi^-\rangle_{f1}\langle \Psi^-|$ and $\hat{\Pi}_{f1}^{({\rm t})}= \hat{I}_{f1} - \hat{\Pi}_{f1}^{({\rm s})}$ are the projector
 operators associated with the singlet and triplet subspaces, respectively, of the $f$-$SQ_1$ system. 
Notice that in the computational basis $\{|\alpha_f \alpha_{1}\rangle\}$ ($\alpha_f,\alpha_1= {\uparrow},{\downarrow}$) a matrix element  $\langle \alpha'_f \alpha'_{1}| \hat{R}_{f1}|\alpha_f \alpha_{1}\rangle$ yields  the probability amplitude that, given the initial joint spin state $|\alpha'_f \alpha'_{1}\rangle$, $f$ is reflected back and the final spin state is $|\alpha_f \alpha_{1}\rangle$ \cite{barrier,CZgate} (an analogous statement holds for $\hat{T}_{f1}$).
Via the identities (\ref{DEFRandT1}) and (\ref{DEFRandT2}) one can easily verify that Eq.~(\ref{defrt})  immediately entails the proper normalization condition $\hat{T}_{f1}^\dag\hat{T}_{f1}+  \hat{R}_{f1}^\dag\hat{R}_{f1}= \hat{I}_{f1}$. Furthermore,
expressed in this form it is now easy to see why the mapping (\ref{mapping1}) is never unitary: in fact for this to happen, $\hat {R}_{f1}$ and  $\hat {T}_{f1}$ should be mutually proportional, i.e.~$r_{{\rm s(t)}}=\xi t_{\rm s(t)}$.
This is impossible since it requires $r_{\rm s}/t_{\rm s}=  r_{\rm t}/t_{\rm t}$, which can be fulfilled only provided that $r_{\rm s}  =r_{\rm t}$ (conflicting with $|r_{\rm s}|  \neq |r_{\rm t}|$ proven above).

\begin{figure}[b]
\begin{center}
\includegraphics[width=0.9\linewidth]{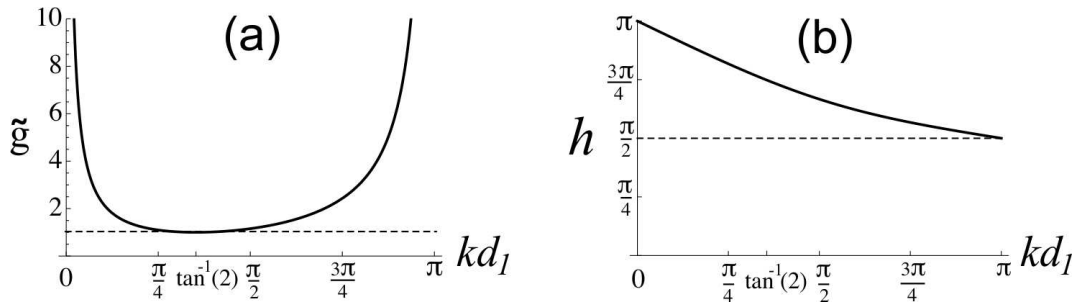}\end{center}
\caption{Plots of the functions $\tilde{g}(kd_1)$ in Eq.~(\ref{g(kd0)}) 
[panel (a)] and $h(kd_1)$ in  Eq.~(\ref{2qstf1}) [panel (b)], which set the conditions for perfect \textsc{swap} between $f$ and the static memories. 
 Either function is periodic of period $\pi$. Note, in particular, that as the optical distance $kd_1$ approaches $n \pi$ ($n=1,2,\ldots$) condition (\ref{g(kd0)}) can be satisfied only in the asymptotic limit of infinite spin-spin coupling. Moreover, there is a threshold $g_{\rm th}=1$ [dashed line in panel (a)] that $g$ must exceed to ensure the existence of values of $kd_1$ allowing for the implementation of the \textsc{swap} gate.}
 \label{fig3}
\end{figure}

A strategy to get around this hindrance is to insert a \textit{perfect mirror} at $x=0$ beyond the SQ located at $x=x_1$ at a distance $d_1$ as sketched in Fig.~\ref{fig2}(a) (this is inspired by \rref\cite{CZgate}, where, however, a somewhat different system was addressed).
First of all, such modified geometry suppresses the transmission channel eliminating the uncertainty
in the direction along which $f$ propagates after interacting with $SQ_1$. 
Specifically, in the presence of the perfect mirror we have $\hat{T}_{f1}^{(\mathrm{m})}= 0$ and
 Eq.~(\ref{mapping1}) thus reduces to
\begin{equation}\rho_{f1} \rightarrow \rho_{f1}'=   \hat{R}^{(\mathrm{m})}_{f1}\rho_{f1}\hat{R}^{(\mathrm{m})\dag}_{f1}, \label{mapping2}
\end{equation}
where now the reflection matrix $\hat{R}_{f1}^{(\mathrm{m})}$ is always unitary $\hat{R}_{f1}^{(\mathrm{m})\dag}\hat{R}_{f1}^{(\mathrm{m})}= \hat{R}_{f1}^{(\mathrm{m})}\hat{R}_{f1}^{(\mathrm{m})\dag}=  \hat{I}_{f1}$.
More interestingly, Eq.~(\ref{mapping2}) allows for the perfect \textsc{swap} gate (\ref{swapgate}) to be implemented.
To see this, observe that since the squared total spin is still a conserved quantity as in  the no-mirror case, the problem reduces to a spinless particle scattering from a \textit{spinless} barrier $\Gamma \delta(x-x_1)$ and a perfect mirror which, via a simple textbook calculation, gives the reflection amplitude 
\begin{equation}
r^{(\mathrm{m})}(\gamma)
=-[i\gamma+(1-i\gamma)e^{2 i kd_1}]/[1+i\gamma(1-e^{2 i k d_1 })]
\end{equation} 
(recall that $\gamma=  \Gamma/k$).
Therefore, a reasoning fully analogous to the previous case leads to 
\begin{equation}
\hat{R}^{(\mathrm{m})}_{f1}= r^{(\mathrm{m})}_{\rm s} \hat{\Pi}_{f1}^{({\rm s})} + r^{(\mathrm{m})}_{\rm t} \hat{\Pi}_{f1}^{({\rm t})}
\end{equation}
with
\begin{eqnarray}
r_{\rm s}^{(\mathrm{m})}&=&  r^{(\mathrm{m})}(-3g/4)   \quad (\textrm{singlet}), \\
r_{\rm t}^{(\mathrm{m})}&=&  r^{(
 \mathrm{m})}(g/4) \qquad  \,(\textrm{triplet}) .
\end{eqnarray}
Observe that $\hat{R}^{(\mathrm{m})}_{f1}$ is unitary because $r^{(\mathrm{m})}(\gamma)$ has unit modulus. 
To work out the conditions for realizing an $f$-$SQ_1$ \textsc{swap} gate (\ref{swapgate}), we use the fact  that this unitary can be written as {$\hat{W}_{f1}=-\hat{\Pi}_{f1}^{({\rm s})} + \hat{\Pi}_{f1}^{({\rm t})}$.
Evidently, $\hat{R}_{f1}^{(\rm m)}$ can be made coincident with $\hat{W}_{f1}$ (up to an irrelevant global phase factor)} if and only if $r_{\rm s}^{(\mathrm{m})}= -r_{\rm t}^{(\mathrm{m})}$. This identity is fulfilled provided that $g$ and $k d_1$ are related to each other according to the function  
\begin{equation}\label{g(kd0)}
g =\tilde{g}(k d_1)=\frac{2}{3}\left(\sqrt{3+4 \cot ^2\!k d_1}-\cot k d_1\right),
\end{equation}
which is plotted in Fig.~\ref{fig3}(a).
Interestingly,  $\tilde{g}(kd_1)\ge1$ means that $g$ must exceed the threshold $g_{\rm th}=1$ to ensure occurrence of the \textsc{swap}.
To summarize, in the presence of a single SQ and for a given spin-spin coupling strength, for any $0<kd_1<\pi$ [see Fig.~\ref{fig3}(a)] there always exists a corresponding coupling constant $G\ge k$ ensuring the occurrence of the $f$-$SQ_1$  \textsc{swap }.
Conversely, as long as $G$ is strictly larger than $k$, there are always two distinct values of $kd_1$ enabling the perfect  \textsc{swap}  between $f$ and $SQ_1$.

Before concluding this section, we point out that, based on the form of $r^{({\rm m})}_{\rm s(t)}$, when the optical distance $kd_1$ is an integer multiple of $\pi$ (i.e.~$k d_1= n \pi$) the above coefficients reduce to $r^{(\mathrm{m})}_{\rm s} = r^{(\mathrm{m})}_{\rm t} =-1$ and hence $\hat{R}^{(\mathrm{m})}_{f1}= - \hat{I}_{f1}$ independently of the
coupling strength. 
This situation is indeed equivalent to \textit{moving the mirror to  $SQ_1$'s location}: the chance for the FQ to be found at such position then vanishes and its spin is thus unable to couple to the SQs. More in general, the property that two objects whose optical separation is an integer multiple of $\pi$ behave \textit{as if they were at the same place} will be exploited repeatedly in this work.

\section{Two static qubits} 
\begin{figure}[b]
\begin{center}
\includegraphics[width=0.45\linewidth]{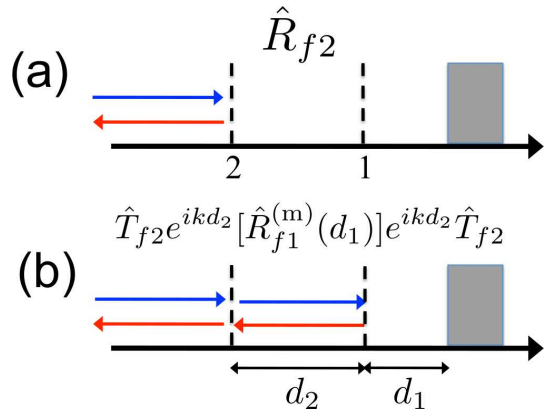}\end{center}
\caption{(Color online) The first-order (a) and the second-order (b) Feynman paths contributing to \eq(\ref{rf12}). }\label{fig4}
\end{figure}

In addition to $SQ_1$ and the perfect mirror, the setup now comprises a further SQ, dubbed $SQ_2$, located on the left of 1 at a distance $d_2$ from it as shown in Fig.~\ref{fig2}(b). Hence, the spin-spin coupling term in $\hat{H}$ now reads 
\begin{equation}\label{v2}
\hat{V}=  G\sum_ {i=1,2}(\hat{\bm \sigma}_f \cdot \bm\hat{\bm \sigma}_{i} )\delta(x-  x_i),
\end{equation}
where $x_1=  -d_1$ and $x_2=  -(d_1+  d_2)$. We aim to implement either an $f$-$SQ_1$ or an $f$-$SQ_2$ \textsc{swap} operation, i.e.~either the unitary $\hat{W}_{f1}\otimes \hat{I}_2$ or $\hat{I}_1 
\otimes \hat{W}_{f2}$, respectively ({note that in any case we require one of the two SQs to be unaffected}).
Analogously to the single-SQ case, the mirror suppresses the transmission channel and thereby one can define a unitary reflection operator $\hat{R}_{f12}$ within the 8-dimensional (8D) overall spin space that fully describes the interaction process output. 
In the spirit of scattering matrices combination via sum over different Feynman paths \cite{datta}, the  scattering operator
  $\hat{R}_{f12}$
   results from a  superposition of all possible paths,  the first of which are sketched in Fig.~\ref{fig4}.
The overall sum is obtained in terms of a geometric series as
\begin{equation}\label{rf12}
\hat{R}_{f12}= \hat{{R}}_{f2}  +  \hat{{T}}_{f2}({\hat{I}_{f12} -  \hat{{R}}_{f1}^{(\rm m)} \hat{{R}}_{f2}}e^{i2 k d_2})^{- 1} \hat{{R}}_{f1}^{(\rm m)}\hat{{T}}_{f2}e^{2 i k d_2},
\end{equation}
where although not shown by our notation, despite it involves qubits $f$ and $SQ_{1(2)}$, each reflection or transmission operator on the right-hand side is intended as the extension to the present 8D spin space.
Also, note that $\hat{{R}}_{f1}^{(\rm m)}$ is a function of $kd_{1}$.

The present setup ensures QST between $f$-{$SQ_1$} and $f$-{$SQ_2$}, respectively, in the regimes
\begin{eqnarray}
& f\mbox{-}SQ_1&\ \ {\rm QST}:\quad kd_2=  h(kd_1),\quad g=\tilde g( k d_1),\label{2qstf1} \\
& f\mbox{-}SQ_2&\ \ {\rm QST}: \quad k d_{1} =  n \pi,\qquad\ \,g=\tilde g( k d_2),\label{2qstf2}
 \end{eqnarray} 
where $n= 1,2,\ldots $, while $h(kd_1)=  \pi- \arg[r^{(\rm m)}_{\rm s}(\tilde{g})]/2$ is a periodic function of period $\pi$ plotted in Fig.~\ref{fig3}(b).
Condition (\ref{2qstf2})  is easily understood: we have already discussed (see the previous section) that when $ k d_1 =  n \pi$ the optical distance between $SQ_1$ and the mirror is effectively zero, hence it is as if the mirror lied at $x=  x_1$ so as to inhibit the $f$-$SQ_1$ coupling. We are thus left basically with the same setup as the one in the previous section, which shows that if condition $g=  \tilde{g}(kd_2)$ is fulfilled  [\cf\eq(\ref{g(kd0)})] then $\hat{R}_{f12}= \hat{I}_1\otimes \hat{W}_{f2}$.

To prove \eq(\ref{2qstf1}), which is key to the central findings in this paper, it is convenient to introduce the coupled spin basis arising from the coupling of $\hat{{\bm \sigma}}_f$, $\hat{{\bm \sigma}}_1$ and $\hat{{\bm \sigma}}_2$. We define $\hat{{\bm S}}_{fi}=  {\hat{\bm \sigma}}_{f}+ \hat{\bm{\sigma}}_{i}$ ($i= 1,2$) and the total spin $\hat{{\bm S}}=  {\hat {\bm \sigma}}_{f}+ \sum_{i=1,2}\hat{\bm{\sigma}}_{i}$. It is then straightforward to check that Eq.~(\ref{v2}) can be expressed as
\begin{equation}
\hat{V}=(G/2)\sum_{i=1,2}(\hat{{\bm S}}_{fi}^2-  3/2)\delta(x-  x_i),\end{equation}
and thus $[\hat{H},\hat{{\bm S}}^2]= 0$ (owing to $[\hat{{\bm S}}_{fi}^2,\hat{{\bm S}}^2]= 0$).
Also, $[\hat{H},\hat{S}_z]= 0$. 
Note, however, that neither $\hat{{\bm S}}_{f1}^2$ nor $\hat{{\bm S}}_{f2}^2$  is conserved since $[\hat{{\bm S}}_{f1}^2,\hat{{\bm S}}_{f2}^2]\neq0$.
Using now the coupling scheme where $\hat{{\bm \sigma}}_f$ is first summed to $\hat{{\bm \sigma}}_1$ \cite{bransden}, the coupled basis reads $\mathcal{B}_{f1}= \{|s_{f1};s,m\rangle\}$, where $s_{f1}$, $s$ and $m=  -s,\ldots,s$ are the quantum numbers associated with $\hat{{\bm S}}_{f1}^2$, $\hat{{\bm S}}^2$ and $\hat{S}_z$, respectively. 
As $s_{f1}= 0,1$ (singlet and triplet, respectively) the possible values for $s$ are $s= 1/2, 3/2$. In the subspace $s= 3/2$ only $s_{f1}= 1$ occurs, while for $s= 1/2$, $s_{f1}$ can be both 0 and 1. 
It should be clear now that given that $s$ and $m$ are good quantum numbers ($\hat{{\bm S}}^2$ and $\hat{S}_z$ are conserved) 
$\hat{R}_{f12}$ is block diagonal in the basis $\mathcal B_{f1}$: four blocks are 1D, each identified by one of the vectors $\{|s_{f1}= 1;s=  3/2,m= -3/2,\ldots,3/2\rangle\}$; two blocks are instead 2D, each spanned by $\{|s_{f1}= 0;s= 1/2,m\rangle,|s_{f1}= 1;s= 1/2,m\rangle\}$ and labeled by $m= -1/2,1/2$.
Due to symmetry reasons, for fixed $s$ the effective form of  $\hat{R}_{f12}$
 in each block is independent of $m$.
Let us first begin with the two $s= 1/2$ blocks.
In the light of the previous section, for both of them, independently of the value of $m$,   we can write  
 $\hat{{R}}_{f1}^{(\rm m)}=  r_{\rm s}^{(\rm m)} |0\rangle\langle 0| + r_{\rm t}^{(\rm m)} |1\rangle\langle 1|$, where we have introduced the concise notation
 $|s_{f1}\rangle= |s_{f1};s= 1/2,m\rangle$. 
As for $\hat{{R}}_{f2}=  \hat{{T}}_{f2}-  \hat{I}_{f2}$, one has to solve an effective scattering problem in a 2D spin space in the presence of the spin-dependent potential barrier $(G/2)(\hat{{\bm S}}_{f2}^2-  3/4-  q_{s_2})\delta(x-  x_2)$, where $s_2$ is the quantum number associated with $\hat{{\bm \sigma}}_2^2$ and we have introduced the discrete function $q_j=  j(j+ 1)$ (here, although $s_2= 1/2$, we leave such quantum number unspecified for reasons that will become clear later on). 
Such task can be carried out easily, as we show in the Appendix. 
Next, by requiring condition (\ref{g(kd0)}),
which ensures that $\hat{{R}}_{f1}^{(\mathrm{m})}$ implements a QST  between $f$ and $SQ_1$ by setting  $r_{\rm s}^{(\rm m)}=-r_{\rm t}^{(\rm m)}$,
and plugging $\hat{{R}}_{f1}^{(\rm m)}$ and $\hat{{R}}_{f2}$ into  \eq(\ref{rf12}), the {matrix elements of ${\hat{R}}_{f12}$} in the $s= 1/2$ block $r_{s'_{f1}s_{f1}}= \langle s'_{f1} |\hat{R}_{f12}|s_{f1}\rangle$ are calculated as 
\begin{eqnarray}
r_{00}&=&{-[}
\tilde{g}^2q_{s_2}
-2  (2-  i \tilde{g}) r_{\rm s}^{(\rm m)}e^{2i k d_2}
-i\tilde{g} (2-iq_{s_2}\tilde{g})r_{\rm s}^{(\rm m)2}e^{4 i k d_2}
]/\Delta,\label{r00}\\
r_{11}&=&{-[}
   i\tilde{g}( 2+iq_{s_2}\tilde{g})
   -2(2+i\tilde{g})r_{\rm s}^{(\rm m)}e^{2 i k d_2}
+q_{s_2} \tilde{g}^2   r_{\rm s}^{(\rm m)2}e^{4 i k d_2}
]/\Delta\label{r11}, \\
r_{01}&=&r_{10}= 2 i\sqrt{q_{s_2}}\,\tilde{g}(1-   r_{\rm s}^{(\rm m)2}e^{4 i k d_2})/\Delta, \label{r01}
\end{eqnarray}
with 
\begin{equation}
\Delta= { -  4}
+i\tilde{g} (1-r_{\rm s}^{(\rm m)}e^{2 i k d_2})
[2+iq_{s_2}\tilde{g}  (1+   r_{\rm s}^{(\rm m)}e^{2 i k d_2})] , 
\end{equation}
(for compactness of notation the dependance of $\tilde g$ on $k d_1$ is not shown).
To realize an $f$-$SQ_1$  \textsc{swap}, i.e.~{$\hat{R}_{f12}=  \hat{I}_2\otimes \hat{W}_{f1}$}, $|s_{f1}= 0\rangle$ and $|s_{f1}= 1\rangle$ must be eigenstates of $\hat{R}_{f12}$ with opposite eigenvalues, namely $r_{00}=  - r_{11}$ must hold. 
Thereby, off-diagonal entries $r_{01}$ must vanish, which yields the condition $r_{\rm s}^{(\rm m)}=  e^{-2i kd_2}$, i.e.~$k d_2=  \pi- \arg [r_{\rm s}^{(\rm m)}(\tilde{g})]/2=  h(kd_1)$\footnote{Strictly speaking, the solution is $k d_2=  n \pi-\arg [r_{\rm s}^{(\rm m)}(\tilde{g})]/2$ for $n=1,2,\ldots$ ($n$ integer). All these solutions are physically equivalent. Lower values of $n$, i.e.~$n\le0$, are to be discarded since they would make $k d_2$ negative.} according to our definition of the $h$ function (see above). 
By replacing this into \eqs(\ref{r00}) and (\ref{r11}) we immediately end up with $r_{00}=  -r_{11}=  1$.

Since for the 1D blocks $s= 3/2$, as mentioned, $s_{f1}$ can only take value 1 and the same occurs for $s_{f2}$ as is easily seen. Hence, $s_{f1}=  s_{f2}= 1$
 and the interaction Hamiltonian is given by  $\hat{V}=  (G/2)\sum_{i=1,2} ( q_{s_{fi}}-  3/2)\delta(x-  x_i)\equiv  (G/4)\sum_{i=1,2}\delta(x-  x_i)$, i.e.~it is effectively spinless.
It should be clear then that the corresponding entry of ${\hat{R}}_{f12}$, denoted by $r^{(3/2)}$, can be found from \eq(\ref{rf12}) through the replacements {${\hat{R}}_{f1}\rightarrow r^{(\rm m)}_{\rm t}$} and ${\hat{R}}_{f2}\rightarrow r_{\rm t}$ (see the previous section). The formerly introduced condition $r_{\rm t}^{(\rm m)}= -r_{\rm s}^{(\rm m)}= -e^{-2i kd_2}$ immediately yields $r^{(3/2)}=  -1$ (matching the value found for $r_{11}$ as it must be given that they both correspond to $s_{f1}= 1$).
This demonstrates that, up to an irrelevant global phase factor, the $f$-$SQ_1$ \textsc{swap} indeed occurs under condition (\ref{2qstf1}). It is important to stress that this result is \textit{independent} of the value taken by $r_{\rm t}$.
In other words, the same result is achieved by replacing $(G/4)\delta(x-  x_2)$ with $\Gamma\delta(x-  x_2)$ with arbitrary $\Gamma$.

\section{Arbitrary number of static qubits} 
We now address the case where an arbitrary number $N$ of SQs are present, the $\nu$th one lying at $x=  x_{\nu}$ in a way that $d_{\nu}=  x_{\nu-1}- {x_{\nu}}$ is the distance between the $\nu$th and $(\nu- 1)$th ones [see Fig.~\ref{fig2}(c)].
Hence, now 
\begin{equation}
\hat{V}=  G \sum_{i=1}^N(\hat{\bm \sigma}_f \cdot \bm\hat{\bm \sigma}_{i})  \delta(x-  x_i).
\end{equation} 
Again, we aim at implementing a selective \textsc{swap} between $f$ and $SQ_\nu$ ($\nu= 1,\ldots,N$). 
Selective QST is achieved for
\begin{eqnarray}\label{eq:2spins_AX1}
\nu<N&:&\quad kd_{i\neq\nu,\nu+ 1}=  n_i\pi,\quad k d_{\nu+ 1}=  h(kd_\nu),\quad g=\tilde g(kd_\nu)\label{alpha},\\
N&:&\quad k d_{i< N} =  n_i \pi,\quad g=\tilde g( k d_N),\label{N}
 \end{eqnarray} 
where $n_i$ can be any positive integer.
Regime (\ref{N}) is immediately explained since it entails that $|x_{N- 1}|$, namely the distance between $SQ_{N-1}$ and the mirror, is a multiple integer of $\pi$, hence the mirror behaves as if it lied at $x=  x_{N- 1}$.
All the static qubits from $SQ_1$ to $SQ_{N-1}$ are thus decoupled from $f$.
We in fact retrieve the case of one SQ at a distance $d_N$ from the mirror, where QST is ensured by condition (\ref{g(kd0)}) (with the replacement $d_1\rightarrow d_N$).

The case in \eq(\ref{alpha}) is explained as follows.
The mirror is effectively positioned at $x=  x_{\nu- 1}$ since each $k d_{i\le\nu- 1}$ is a multiple integer of $\pi$. 
On the other hand, $k d_{i>\nu+ 1}= n_i \pi$ holds as well: the static qubits indexed by $i$ such that $\nu+ 1\le i\le N$ behave as if they were \textit{all} located at $x=  x_{\nu+ 1}$.
Thereby, effectively $\hat{V}=  G\sum_{i=\nu+1}^N(\bm\hat{\bm \sigma}_f\cdot\bm\hat{\bm \sigma}_{i})\delta(x-  x_{\nu+1})+  G(\bm\hat{\bm \sigma}_f\cdot\bm\hat{\bm \sigma}_{\nu}) \delta(x-  x_{\nu})$ (subject to a hard-wall boundary condition at $x=  x_{\nu-1})$.
Let $\bm\hat{\bm \sigma}_{\rm eff}= \sum_{i=\nu+1}^N\bm\hat{\bm \sigma}_{i}$ be the total spin of the $N- \nu$ SQs effectively located at $x=  x_{\nu+1}$ and $s_{\rm eff}$ the quantum number associated with $\bm\hat{\bm \sigma}_{\rm eff}^2$.
For {$N-\nu$} even, $s_{\rm eff}= 0,1,\ldots,(N- \nu)/2$, while for {$N-\nu$} odd $s_{\rm eff}= 1/2,3/2,\ldots,(N- \nu)/2$.
As, clearly, $s_{\rm eff}$ is a good quantum number, in each subspace of fixed $s_{\rm eff}$ an effective static spin-$s_{\rm eff}$ particle lies at $x=  x_{\nu+1}$\footnote{Unlike a very spin-$s_{\rm eff}$ particle, in our case a given value of $s_{\rm eff}$ can exhibit degeneracies (\eg for $N= 3$ the value $s_{\rm eff}= 1/2$ is two-fold degenerate). Yet, such degeneracies do not play any role here and can in fact be ignored.}.
By coupling this spin to $f$ and $SQ_\nu$, we find that the total quantum number can take values $s=  s_{\rm eff}- 1,s_{\rm eff}, s_{\rm eff}+ 1$ (we can assume $s_{\rm eff}\ge1$ since the case $s_{\rm eff}=1/2$ has been analyzed in the previous section).
Among these, only $s=  s_{\rm eff}$ is degenerate since in the corresponding eigenspace either $\hat{\bm S}_{f\nu}^2$ or $\hat{\bm S}_{f\rm e}^2=(\bm\hat{\bm \sigma}_f+\bm\hat{\bm \sigma}_{\rm eff})^2$ can take \textit{two} possible values, i.e.~$s_{f\nu}=  0,1$ and $s_{f\rm e}=  s_{\rm eff}\pm1/2$ ($s_{fe}$ is the quantum number associated with $\hat{\bm S}_{f\rm e}^2$).
The reflection matrix for the system is thus block-diagonal, where each block corresponding to either $s= s_{\rm eff}-1$ or $s= s_{\rm eff}+1$ is 1D, while a block corresponding to $s=s_{\rm eff}$ is 2D\@.
In the latter case, the corresponding reflection amplitudes in the basis $\{|s_{f1}; s,m_s\rangle=  |s_{f1}\rangle\}$ can then be worked out in full analogy with the $s= 1/2$ subspace in the case of two SQs (see the previous section).
Hence, they are given by Eqs.~(\ref{r00})--(\ref{r01}) under the simple replacements $s_2\rightarrow s_{\rm eff}$, $d_1\rightarrow d_{\nu}$ and $d_2\rightarrow d_{\nu+1}$.
Thereby, $f$-$\nu$ QST occurs for $r_{t}^{(\rm m)}=-  r_{s}^{(\rm m)}=  -  e^{-2i kd_{\nu+ 1}}$, which holds provided that $g= \tilde{g}(kd_\nu)$ and $kd_{\nu+1}=  h(kd_\nu)$. 
On the other hand, for $s=  s_{\rm eff}-1$ ($s=  s_{\rm eff}+1$)   we have  $s_{f\rm e}=  s_{\rm eff}- 1/2$ ($s_{f\rm e}=  s_{\rm eff}+1/2$), while $s_{f\nu}= 1$.
Hence, similarly to the $s= 3/2$ case in the previous section, in either of these subspaces the interaction Hamiltonian has the spinless effective form $\hat{V}=  (G/2)( q_{s_{\rm eff}\pm1/2}- 3/4-  q_{s_{\rm eff}}) \delta(x-  x_{\nu+1})+  (G/4)\delta(x-  x_{\nu})$. The condition $r_{t}^{(\rm m)}= -  r_{s}^{(\rm m)}=  -  e^{-2i kd_{\nu+1}}$ then ensures that in each case the corresponding overall reflection amplitude equals $-1$ (see the comment at the end of the previous section).
A \textsc{swap} operation between $f$ and $SQ_\nu$ is therefore implemented.

\section{Working conditions} 
Based on the above findings, in particular \eq(\ref{alpha}), the following working conditions to achieve selective writing/read-out of the static register can be devised. 
Firstly,  one fixes once for all the desired coupling strength $g= g_0$ [provided that it exceeds the threshold value $g_{\rm th} =1$, equivalent to $G = k$; see Fig.~\ref{fig3}(a)].
Next, we  choose one of the two different distances (in unit of $k^{-1}$) that correspond to $g= g_0$ according to the function $\tilde g(kd)$ [see Fig.~\ref{fig3}(a)].
Let us call such a distance $d_a$, which therefore fulfills  $\tilde g(kd_a)\equiv g_0$.
A further distance $d_b = h(kd_a)/k$ [\cf Fig.~\ref{fig3}(b)] is then univocally identified.
All the nearest-neighbour distances are set equal to an integer multiple of $\pi$ (in unit of $k^{-1}$) but the $\nu$th and $(\nu+1)$th ones, which are set to $d_a$ and $d_b$, respectively. {In a practical implementation, such tunable setting of nearest-neighbor distances could be achieved by fabricating the setup in such a way that the FQ can propagate along three possible paths instead of a single one (similarly to the geometry of the well-known Aharonov-Bohm rings). If the paths have different lengths, the actual path followed by the FQ can be chosen {by means of tunable beam splitters}, in fact setting the effective SQ-SQ distance.}

{In practice, unavoidable static disorder will affect the ideal pattern of nearest-neighbor SQ distances.  Through a proof-of-principle resilience analysis 
we have assessed that, by assuming Gaussian noise and in the case of a single SQ, an uncertainty in its position of order of about $10\%$ yields a process fidelity above the 95\%-threshold. This witnesses an excellent level of tolerance, in line with similar tests \cite{CZgate, ps}. Preliminary studies for the cases of two and three SQs have been carried out as well, confirming comparable performances. A comprehensive conclusive characterization of the effects of static disorder in the case of an arbitrary number of SQs, though, requires a rather involved analysis and thus goes beyond the scopes of this paper.}

\section{Conclusions}
We have considered a typical scenario envisaged in distributed quantum information, where writing and read-out of a register of SQs is performed through a FQ\@.
In a fully quantum theory, the motional degrees of freedom (MDOFs) of the FQ should be treated as quantum, which is expected to substantially complicate the dynamics.
By taking a paradigmatic Hamiltonian, we have discovered that, as long as the $f$-$SQ$ coupling is above a certain threshold value (i.e.~$G\geq k$ with $k$ being the input momentum of the FQ), for an arbitrary number of SQs \textit{selective} QST can be achieved on demand by tuning only two SQ distances.

Throughout, as is customary in scattering-based theories, we have assumed to deal with a perfectly monochromatic plane wave for the flying qubit.
In practice, clearly, this is a narrow-bandwidth wavepacket centered at a carrier wave vector $k_0$.
A detailed resilience study of the performances of our protocol in such conditions is beyond the scope of the present paper.
Yet, similarly to \rrefs\cite{CZgate,ps,nops}, it is reasonable to expect the gate fidelity to be only mildly affected owing to the smoothness of functions $\tilde g(kd)$ and $h(kd)$ (\cf Fig.~\ref{fig3}).
In our model, we assumed a Heisenberg-type spin-spin interaction.
As already stressed, our attitude here was to take this well-known coupling as a paradigmatic model to show the possibility that selective writing/read-out is in principle achievable. 
{However, there exist setups where the Heisenberg-type coupling occurs so as to make them potential candidates for realizing our protocol. For instance (see also {Refs.~}\cite{jeff}) this is the case of an electron propagating along a semiconducting carbon nanotube \cite{tans} {and scattered from} single-electron quantum dots or molecular spin systems featuring unpaired electrons, such as Sc@C82 \cite{morton}.}
Alternatively, one can envisage a photon propagating in a 1D waveguide to embody the FQ in a way that its spin is encoded in the polarization DOFs.
A three-level $\Lambda$-type atom could then work as the static qubit, where the $\{|{\uparrow\rangle,|{\downarrow}\rangle}\}$ basis is encoded in the ground doublet, while each transition to the excited state requires orthogonal photonic polarizations;  see Refs.~\cite{law,ref:AtomPhoton}.
Although similar, the corresponding (pseudo) spin-spin coupling, yet, is not equivalent to a Heisenberg-type one.
We found some numerical evidence that this alternative coupling model could work as well, at least in the few-SQ case.
An analytical treatment, however, is quite involved and thus no definite answer can be given.
This is connected to the question whether some specific symmetry is a necessary prerequisite for such remarkable effects to take place (in passing, note that the Heisenberg model conserves the squared total spin, which was crucial to carry out our proofs).
All these issues are the focus of ongoing investigations.

It is worth mentioning that in a recent work \cite{lovett}, Ping \textit{et~al.} proposed a protocol for imprinting the quantum state of a ``writing" FQ on an array of SQs and retrieving it through a ``reading" FQ at a next stage  \cite{lovett}.
There, information is intentionally encoded over the \textit{entire} register, which has some advantages, while MDOFs are in fact treated as classical.
Significantly enough, here we have shown that the inclusion of quantum MDOFs can allow for control over \textit{local} encoding/decoding.
In line with other works \cite{nops}, such apparent complication appears instead a powerful resource to carry out refined QIP tasks.

\section*{Acknowledgements}
We thank G. Cordourier-Maruri, S. Bose, D. Browne, S. Lloyd and H. Schomerus for useful discussions.
We acknowledge support from MIUR through the FIRB-IDEAS project RBID08B3FM, the Italian Ministry of Foreign Affairs through the Joint Italian-Japanese Laboratory on Quantum Technologies: Information, Communication and Computation and the  Ministry of Education, Culture, Sports, Science and Technology, Japan, through Grant-in-Aid for Young Scientists (B) (No.~21740294). KY is also supported by a Waseda University Grant for Special Research Projects (2012A-878).

\section*{Appendix. Derivation of ${\hat R}_{f2}$ in the basis $\{|s_{f1}\rangle\}$}
Here, we derive the matrix elements of the operator ${\hat{R}}_{f2}$ in the degenerate subspace $s=  s_2$ (which is $s= 1/2$ in the case of $N= 2$ SQs), namely all the reflection coefficients $\bar r_{s'_{f1}s_{f1}}= \langle s'_{f1} |\hat{R}_{f2}|s_{f1}\rangle$ in terms of the basis $\mathcal{B}_{f1}= \{|s_{f1}= 0,1\rangle\}$, where $|s_{f1}\rangle= |s_{f1};s=  s_2,m\rangle$.
In line with the main text, we give the proof without specifying $s_2$ (which can thus be any positive integer or semi-integer number). 
The Hamiltonian reads $\hat{H}= {\hat{p}^2}/{2} + \hat{V}$ with $\hat{V}=  (G/2)(\hat{{\bm S}}_{f2}^2- 3/4-  q_{s_{2}})\delta(x)$ (we have set $x_2= 0$ since the result is evidently independent of $x_2$).
The key task is to work out the matrix representation of $\hat{{\bm S}}_{f2}^2$ in the basis of eigenstates of $\hat{{\bm S}}_{f1}^2$, $\mathcal{B}_{f1}= \{|s_{f1}= 0,1\rangle\}$.
We first observe that in the present $s=  s_2$ subspace $s_{f2}=  s_2\pm1/2$. 
Accordingly, the scheme where $f$ is first coupled to 2 leads to the alternative basis $\mathcal{B}_{f2}= \{|s_{f2}=  s_2\pm1/2\rangle\}$ such that {$\hat{S}^2_{f2}|s_{f2}=  s_2\pm1/2\rangle=  q_{s_{f2}}|s_{f2}=  s_2\pm1/2\rangle$}.
Thereby, in the basis $\mathcal B_{f2}$, $\hat{{\bm S}}_{f2}^2$ has the diagonal matrix representation $\diag(q_{s_2- 1/2},q_{s_2+ 1/2})$.
The transformation matrix between the two basis can be calculated through 6$j$ coefficients \cite{bransden} as
\begin{equation}
\langle s_{f2}|s_{f1}\rangle= (-1)^{s_2+1}\sqrt{(2 s_{f1}+1)(2 s_{f2}+1)}
\left\{
\begin{array}{ccc}
  s_2&           1/2&           s_{f2}         \\
            1/2          &s_2&           s_{f1}\end{array} \right\}.
\end{equation}
Using these then yields $\hat{{\bm S}}_{f2}^2$ in the basis $\mathcal B_{f1}$ as
\begin{eqnarray}
\langle 0|\hat{{\bm S}}_{f2}^2|0\rangle=  -\frac{3}{8}+ \frac{q_{s_2}}{2},
\qquad
\label{s00}
\langle 1|\hat{{\bm S}}_{f2}^2|1\rangle= -\frac{7}{8}+ \frac{q_{s_2}}{2},\label{s11}\\
\langle 0|\hat{{\bm S}}_{f2}^2|1\rangle=  \langle 1|\hat{{\bm S}}_{f2}^2|0\rangle= \frac{\sqrt{q_{s_2}}}{2}.\label{s01}
\end{eqnarray}
Next, in close analogy with \rrefs\cite{NJP1,JPA}, we search for a stationary state $|\Psi_{s'_{f1}}\rangle=  \varphi_{s_{f1}'0}(x)|0\rangle+  \varphi_{s_{f1}'1}(x)|1\rangle$ such that $\hat{H}|\Psi_{s'_{f1}}\rangle=   ( k^2/2 )  |\Psi_{s'_{f1}}\rangle$, where $s_{f1}'= 0,1$ labels the initial spin state (prior to the interaction process). 
Each  function $\varphi$  has the form
\begin{equation}\label{ansatz}
\varphi_{s_{f1}'s_{f1}}(x)=  (\delta_{s_{f1}'s_{f1}}e^{i kx}+  \bar{r}_{s_{f1}'s_{f1}}e^{-i kx})\theta(-x) +  \bar t_{s_{f1}'s_{f1}}e^{i kx}\theta(x).
\end{equation}
The unknown coefficients, including $\{\bar r_{s_{f1}'s_{f1}}\}$, i.e.~the entries of ${\hat{R}}_{f2}$, can be found by imposing the continuity condition of $\varphi_{s_{f1}'0}(x)$ and $\varphi_{s_{f1}'1}(x)$ at $x=  0$ and the two constraints
\begin{eqnarray}
\Delta \varphi_{s_{f1}',0}'(0)&= & G {\sqrt{q_{s_2}}}\,\varphi_{s_{f1}',1}(0),\label{boundcondacc1}\\
\Delta \varphi_{s_{f2}',1}'(0)&= &{-G}\varphi_{s_{f1}',1}(0) +  G {\sqrt{q_{s_2}}}\,\varphi_{s_{f1}',0}(0),\label{boundcondacc2}
\end{eqnarray}
where $\Delta\varphi'_{s_{f1}',s_{f1}}(0)$ is the jump of the derivative at $x=  0$. 
With the help of \eqs(\ref{s00}) and (\ref{s01}), \eqs(\ref{boundcondacc1}) and (\ref{boundcondacc2}) can be straightforwardly obtained from the Schr\"{o}dinger equation by integrating it across $x=  0$ and then projecting onto $|0\rangle$ and $|1\rangle$ \cite{JPA}.
By solving the linear system in the cases $s_{f1}'= 0,1$, we thus end up with  
\begin{eqnarray}
\bar r_{00}=\langle 0|\hat{R}_{f2}|0\rangle=q_{s_2}g^2/\Delta_{s_2},\qquad
\bar r_{11}= {-ig}(2+iq_{s_2}g)/\Delta_{s_2} ,\\
\bar r_{01}=\langle 0|\hat{R}_{f2}|1\rangle=\langle 1|\hat{R}_{f2}|0\rangle^*
=2 i\sqrt{q_{s_2}}\, g  /\Delta_{s_2} , 
\end{eqnarray}
where $\Delta_{s_2}= -4+ 2 i g -q_{s_2}g^2$.

\end{document}